\documentclass[11pt]{article}
\usepackage[colorlinks=true,citecolor=blue,linkcolor=blue]{hyperref}
\usepackage{multirow}
\usepackage[left=2cm,right=2cm,top=2cm,bottom=2cm]{geometry}
\usepackage{cite}
\usepackage{psfrag}
\usepackage{graphicx}
\usepackage{epsfig}
\usepackage{booktabs}
\usepackage{amsmath,amsfonts,amssymb}
\usepackage{pstcol,pst-fill,pst-grad}
\usepackage{pstricks,pst-fill,pst-grad}
\usepackage{euscript}
\usepackage{pstricks}
\usepackage{wrapfig}
\usepackage{slashed}
\usepackage[toc,page]{appendix}
\usepackage{here}
\usepackage{multirow}
\usepackage{ulem}

\date{}
\begin{document}
	\title{\vspace{-3cm}
		\hfill\parbox{4cm}{\normalsize \emph{}}\\
		\vspace{1cm}
		{Theoretical investigation of the physical properties of cubic perovskite oxides $SrXO_3 \;(X =Sc,\; Ge,\; Si)$
	}}
	\vspace{2cm}
	
	\author{A Waqdim$^{1}$, M Agouri$^{1}$, A Abbassi$^{1,}$\thanks{Corresponding author, E-mail: abbassi.abder@gmail.com},
		B. Elhadadi$^{1}$, Z. Zidane$^{1}$, S Taj$^{1}$, B Manaut$^{1,}$\thanks{Corresponding author, E-mail: b.manaut@usms.ma},\\ M Driouich$^1$ and M El Idrissi $^{1,2}$ \\\\
		{\it {\small$^1$ Laboratory of Research in Physics and Engineering Sciences,}}\\
		{\it{\small Sultan Moulay Slimane University, Polydisciplinary Faculty, Beni Mellal, 23000, Morocco.}}\\
		{\it {\small$^2$ Sultan Moulay Slimane University, Polydisciplinary Faculty of Khouribga.}}
	}	
	\maketitle \setcounter{page}{1}
	\date{\today}
	
	\begin{abstract}
		Various physical properties (electronic, optical and thermoelectric) of cubic perovskite oxides $SrXO_3\; (X=Sc,\; Ge,\; Si)$ are investigated by using the density functional theory (DFT) within Wien2k code. This code is based on different approximations such as generalized gradient approximation GGA, PBEsol, LDA, WC and the modified Becke-Johnson exchange potentials (mBJ, nmBJ and unmBJ). 
		Structural properties and the optimization have been calculated using PBEsol functional which showed a significant results that are in good agreement with the experimental ones. The results for physical properties such as electronic, thermoelectric and optical are analyzed in detail by using the nmBJ approximation. The obtained results present an opening gap for $SrSiO_3$, $SrGeO_3$ and a metallic behavior for $SrScO_3$. An average of transmittance which is about $94\%$ to $97\%$ was observed in the range of visible light. The increase of electrical conductivity with temperature confirms the effect of thermal agitation on the concentration of charges carriers using BolzTrap Package. Lastly a phonon dispersion was made and show that for $X=\; Ge$ and $Si$, the structure is relatively stable while $SrScO_3$ is dynamically unstable. These results prove the ability that these materials can be exploited in many applications and manufacturing of optoelectronics for Sensors.
	\end{abstract}
	Keywords:   DFT, Perovskite, wien2K, nmBJ, BolzTrap.
	
	\maketitle
	\section{Introduction}
	Oxides-Perovskite $ABO_3$ is being recognized among the most promising materials with wide applicability in clean energy applications. This came mainly from the quality presented by these materials as well as their simple manipulation in order to change the nature of $A$ and $B$ presented in there phases. The modification of these elements and their substitutions by others leads to  make a deep change in their intrinsic properties which opens a huge field and research discipline in order to study their physical properties according to the chemical and electronic nature of  the atoms $A$ and $B$.\\
	Calcium Titanate was the first perovskite materials investigated which was discovered since many years by German mineralogist and crystallographer Gustau Rose, who determined its physical properties and chemical composition, and named after by the Russian mineralogist Alekseivitch Perovski \cite{1}. The general chemical formula of perovskite materials was $ABX_3$, where $A$ and $B$ are two cations ($A$ Larger than $B$) and $X$ is an anion. Acutely Perovskite designates a family of materials which all have the same atomic arrangement. The anions in their phase can be an oxide, sulphide and, bromide, chloride, fluoride or hydride. The perovskite phases occupy a great place in the ternary systems known under the $ABX_3$ composition with $X$ is taken by Oxygen. $ABO_3$ consisted then of three-dimensional chain $BO_6$ octahedra which are connected by the vertices while the $A$ cation is surrounded by 12 oxygen atoms. They received a great interest for their simple cubic crystal structure and diverse physical and chemical properties, that made them suitable for many interesting technologies by exploiting their superconductivity in microelectronics \cite{2}, solar cells, optoelectronic feature \cite{3,4}, and suitable also in several advanced applications such as piezoelectricity, magneto-transport, thermoelectric, giant magneto-resistance, charge ordering, ferroelectricity \cite{5, 6, 7}.\\
	Many perovskites oxides are cubic or closely cubic, however, they often undergo one or more structural phase transitions, particularly at low temperature \cite{8} such as $CsPbBr_3$ which presents an orthorhombic at 361 $K$ and tetragonal at 361 $K < T < 403 K$ and then cubic at 403 $K$ \cite{9}. The $BaTiO_3$ undergoes also a variation phase which is shown tetragonal phase at 300 $K$ and cubic phase at 450 $K$ \cite{10}. Furthermore the ideal structure of Perovskite oxides in term of structural stability is a cubic with lattice parameters in the range 3,6-5 $\AA$  and space group Pm-3m.\\
	Cubic $SrXO_3\; (X=Sc,\; Ge,\; Si)$ belong to the perovskites family which not widely reported experimentally and theoretically. Some experimental studies recently published confirm that $SrGeO_3$ is the cubic structure with lattice parameters $a = b = c = 3.798\; \AA$ \cite{11, 12} using the high-resolution X-ray diffraction (HR-XRD).\\
	Others showed that $SrGeO_3$ has excellent conductivity and stability at low temperatures \cite{13}, and has a good application prospect in the field of optics and autocatalysis due to its wide band gap of $3.2 \;eV$. The $SrGeO_3$ and $SrSiO_3$ were also treated in order to exploit and expand their use in other applications such as electrolyte materials in solid oxide fuel cells (SOFC) \cite{14, 15}. Some other studies reported the application of different morphology of $SrSiO_3$ in the photocatalytic decomposition combined with a high reactivity of reactants, having low energy consumption and less air and water pollution \cite{16}. Theoretically, thermoelectric, optoelectronics, thermodynamic and magnetic properties of cubic perovskite $SrMnO_3$, was investigated \cite{17} which could be used in different optical devices for visible and ultraviolet light. Recent works reported and interested in calculation of the structural, electronic, magnetic and elastic properties of the transition metals based perovskites $SrTMO_3\; (TM = Mn,\; Fe,\; Co,\; Tc,\; Ru,\; Rh,\; Re,\; Os$ and $Ir)$ and cubic $SrAO_3\;(A = Ca,\; Ge)$ in order to show their physical stability Ferro/Antiferromagnetic and thermoelectric properties \cite{18, 19}.
	
	The investigation of the physical properties of these perovskites is important to understand more their behaviors and expand their exploitation in broad spectrum such as in optoelectronic devices like solar cell, sensors, diode, laser, etc. In this perspective, the aim of the present work is to investigate structural, electronic, optical and thermoelectric properties of cubic $SrXO_3\; (X=Sr,\; Ge,\; Si)$. To do that, we used the new-mBJ approximation due to its calculation reliability. We used also Boltzman’s semi-classical equation to study the influence of the temperature on the transport phenomena such as electrical and thermal conductivities, Seebeck coefficient and ZT parameter.
	\section{Materials and Methods}
	The investigations of the electronic, optical and transport properties for the cubic perovskites $SrGeO_3$, $SrSiO_3$ and $SrScO_3$ were calculated using the full potential linearized augmented plane wave plus local orbitals (FP-LAPW + lo) method based on the Density Function Theory (DFT) \cite{20, 21} developed in the WIEN2k code \cite{22}. Many research shows that DFT can evaluate an exact exchange energy, the limitation of this method is not in the term of exchange but in correlation functional that is still unknown, therefore the use of approximations is needed. Most of the time the correlation functional is considered with a really small effect and the approximations are significant\cite{23}. For structural optimization properties, we utilized several exchange correlation potentials which are Local Density Approximation (LDA) \cite{24}, Perdew-Burke-Ernzerhof Generalized Gradient Approximation (PBE-GGA) \cite{25}, Wu-Cohen GGA (WC-GGA) \cite{26} and Perdew-Burke-Emzerhof for solids GGA (PBEsol-GGA) \cite{27}.\\	
	To investigate other properties, we used the modified Becke Johnson GGA (mBJ- GGA) \cite{28}, new mBJ GGA (nmBJ-GGA) \cite{29} and the unmodified mBJ GGA (unmBJ-GGA) methods, which are given by:
	\begin{equation}
		\vartheta^{mBJ}_{x, \sigma}(r)=c \vartheta^{BR}_{x, \sigma}(r)+(3c-2)\dfrac{1}{\pi}\sqrt{\dfrac{5}{12}}\sqrt{\dfrac{2t_{\sigma}(r)}{\rho_{\sigma}(r)}},
	\end{equation}
	The Becke and Johnson (BJ) devlopped an exact exchange potential without any empirical parameter, it can be write as follow:
	\begin{equation}
		\vartheta^{BJ}_{x, \sigma}(r)= \vartheta^{BR}_{x, \sigma}(r)+\dfrac{1}{\pi}\sqrt{\dfrac{5}{12}}\sqrt{\dfrac{2t_{\sigma}(r)}{\rho_{\sigma}(r)}},
	\end{equation}
	where $\rho_{\sigma}=\sum_{i=1}^{N_{\sigma}}\mid\Psi_{i,\sigma}\mid^{2} $  is the electron density, $t_{\sigma}=(\dfrac{1}{2})\sum_{i=1}^{N_{\sigma}}\nabla\Psi_{i,\sigma}^{*} \nabla\Psi_{i,\sigma}$ is the kinetic-energy density. \\
	And \hspace{2cm} $\vartheta^{BR}_{x, \sigma}(r)=-\dfrac{1}{b_{\sigma}(r)}\left( 1-e^{({-x_{\sigma}(r)})}-\dfrac{1}{2}x_{\sigma}(r)e^{({-x_{\sigma}(r)})}~\right)$\\
	is the Becke-Roussel (BR) exchange potential. $x$ is determined from a nonlinear formalism, where $\rho$, $\nabla \rho$, $\nabla\rho^{2}$, $t$, and then $b$ is calculated with : 
	\begin{equation}
		b = [x^{3} e^{-x}/(8 \pi \rho)]^{1/3}
	\end{equation}
	\\ and $c$ is given by: 
	\begin{equation}
		c=A+B\sqrt{\overline{g}}
	\end{equation}
	where \hspace{2cm}
	$\overline{g}= \dfrac{1}{V_{cell}}\int_{cell}\dfrac{1}{2}\left( \dfrac{\mid\nabla\rho^{\uparrow}(r)\mid}{\rho_{\uparrow}}\dfrac{\mid\nabla\rho^{\downarrow}(r)\mid}{\rho_{\downarrow}}\right) dr^{3}$\\
	\vspace{0.5cm}
	is the average of $g = \mid\nabla\rho\mid / \rho$ in the unit cell of considered volume $V_{cell}$. $A$ and $B$ are parameters which are equal $A = -0.012$ and $B = 1.023\; bohr^{1/2}$.\\
	Depending on the experimental band gaps fit, the $c$ values larger than $1$ give to a less negative potential, especially for low-density regions.
	Static calculations are made using 1000 k-points in the reciprocal space and under the energy convergence limit of $10^{-5} \;Ry$. The multiplication of the smallest of all atomic sphere radii ($R_{MT}$) and the maximum modulus of the reciprocal vector ($K_{max}$) is estimated to be equal to 7 ($R_{MT} \times K_{max} = 7$). The Muffin tin radii ($R_{MT}$ ) used are $(2.4,\; 1.84,\; 1.61)$ for $SrGeO_3$, $(2.31, \;1.7,\; 1.6)$ for $SrSiO_3$ and $(2.5, \;1.95, \;1.75)$ for $SrScO_3$, respectively. Thermoelectric properties for the $SrXO_3\; (X = Sc, Ge, Si)$ have also been investigated using BoltzTraP package \cite{30}.
	\section{Results and discussion}\label{Sec2}
	\subsection{Structural and Electronic properties}
	
\begin{figure}[H]
		\centering
		\includegraphics[scale=0.4]{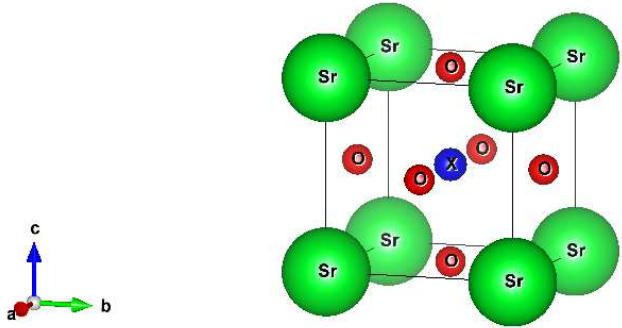}
		\caption{Perovskite cubic  structure of $SrXO_3\; (X = Sc,\;Ge,\;Si)$ \cite{31}.}
		\label{Figure:1}
	\end{figure}

\begin{figure}[H]
		\centering
		\includegraphics[width=1\textwidth,height=7 in]{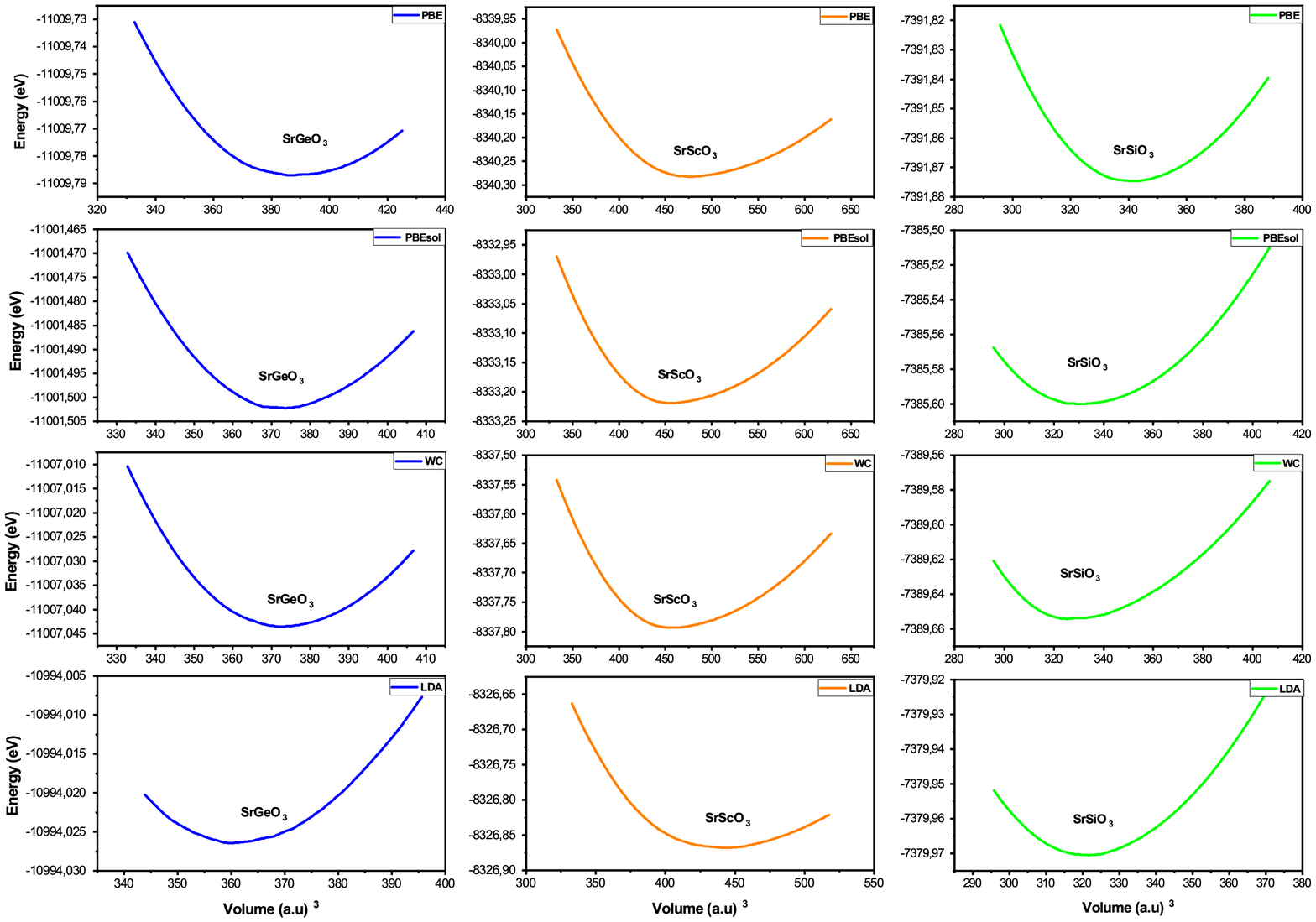}
		\vspace{-2.2cm} 
		\caption{Total energy (Ry) versus total volume $ (a.u^{3} ) $ for cubic perovskites $SrXO_{3} \; (X=Si,\;Ge,\;Sc)$ calculated for LDA, WC-GGA, PBE-GGA and PBEsoL-GGA methods.}
		\label{Figure:2}
	\end{figure}

	The perovskite oxides $SrXO_3 \;(X=Sc,\; Ge,\; Si)$ possess a cubic crystal structure with the space group Pm-3m (no.221) and the experiment lattice parameter $a = 3.798\AA$ \cite{12}. Figure \ref{Figure:1} shows that the unit cell of the perovskite contains five atoms in a single formula unit. In the unit cell, the $Sr$ atoms occupy the corner positions with $1a$ Wyckoff site and $(0, 0, 0)$ fractional coordinates, the $X$ atom occupies the centred position with $1b$ Wyckoff site and $(1/2, 1/2, 1/2)$ fractional coordinates and the $O$ atoms occupy the face centred positions with $3c$ Wyckoff sites and $(0, 1/2, 1/2)$ fractional coordinates.
	The structural properties such as lattice constant ($a$), volume ($V_0$), hydrostatic bulk modulus ($B_0$) and its pressure derivative  $B'$ were obtained by using the equation of states according to Birch-Murnaghan \cite{32}:
	\begin{equation}
		E(V)= E_{min} +\dfrac{9V_{0}B_{0}}{16} \left\lbrace{\left\lbrace \left(\dfrac{V_{0}}{V}\right) ^{2/3} - 1\right\rbrace}^{2} B^{'} + {\left\lbrace \left( \dfrac{V_{0}}{V}\right) ^{2/3} - 1\right\rbrace}^{2}{\left\lbrace 6 - 4\left( \dfrac{V_{0}}{V}\right) ^{2/3}\right\rbrace}\right\rbrace \label{1}
	\end{equation}
	Where $E_{min}$ and $V_0$ are the equilibrium energy and volume. Based on the equation (\ref{1}) and by using various methods (LDA, WC-GGA, PBE-GGA and PBEsoL-GGA), we have obtained the total electronic energy ($E$) with respect to the unit cell volume ($V$) as shown in figure \ref{Figure:2}.
The results of structural properties such as lattice constants $a$, equilibrium volume $V_0$ , bulk modulus $B_0$ and derivative of bulk modulus $B'$ using different approximations are summarized in table \ref{Table:1}. 
For LDA method, It is well known to have some problems such as overbinding \cite{33}. The PBE-GGA was introduced to address and improve some of the weaknesses of the LDA \cite{34}, in particular, it frequently overcorrects LDA’s overbinding. In addition, PBEsol-GGA and WC-GGA approximations decreased such predicaments \cite{27} and they were given much better lattice parameters than PBE-GGA.
\begin{table}[h]	
	\begin{center}
		\begin{tabular}{ lllllll }
			\hline
			Compounds   & Méthods    & $a$(\AA)& $V_{0}$(\AA$^{3})$  & $B_{0}(GPa)$  & $B'$ \\ 
			\hline
			\multirow{4}{*}{SrScO$_3$}  
			&LDA        &4.0325      &65.5727  &139.8641  &3.8889        \\
			&PBE-GGA    &4.1316      &70.5269  &116.2384  &4.0589		 \\
			&PBEsol-GGA &4.0772      &67.7775  &126.4539  &4.0790        \\
			&WC-GGA     &4.0770      &67.7676  &126.8915  &4.1037        \\
			&Others\cite{35}&4.140    &70.9579  &          &             \\
			\hline
			\multirow{4}{*}{SrGeO$_3$}  
			&LDA        &3.7684      &53.5179  &199.1564  &4.7856        \\
			&PBE-GGA    &3.8574      &57.3963  &157.0127  &5.144         \\
			&PBEsol-GGA &3.8082      &55.228   &179.5736  &4.6567        \\
			&WC-GGA     &3.8101      &55.3107  &179.8121  &4.4057        \\
			&Exp\cite{12}&3.798      &54.7854  &          &              \\		
			&Others\cite{19}&3.807   &55.1758  &178.25    &              \\		
			\hline
			\multirow{4}{*}{SrSiO$_3$}   
			& LDA       &3.6234      &47.5717  &237.7161  &4.4355        \\
			& PBE-GGA   &3.6973      &57.3963  &200.4158  &4.4802        \\
			& PBEsol-GGA&3.6577      &48.9355  &218.6773  &4.4643        \\
			& WC-GGA     &3.6541     &48.7911  &220.3336  &6.5902        \\
			& Exp\cite{36}&3.664     &49.18    &211       &4.0000        \\
			\hline
		\end{tabular}
	\end{center}
	\caption {Calculated lattice constants ($a$), equilibrium volume ($V_{0}$), bulk modulus ($B_{0}$) and pressure derivative of bulk modulus ($ B' $) for $SrXO_{3}  \;(X=Sc,\;Ge,\;Si)$ cubic perovskite compounds.}
	\label{Table:1}
\end{table}\\
Revised density-functional PBEsol improves predictions of equilibrium properties of solids. It was very recently proposed by J.P. Perdew \textit{et al.} \cite{27} and reported by other works \cite{37, 38}. From table \ref{Table:1}, It has been seen that PBEsol method showed good agreement with experimental and theoretical works \cite{12,35,36}. For all these reasons, structural optimizations for our systems were studied by using the PBEsol method.

\begin{figure}[H]
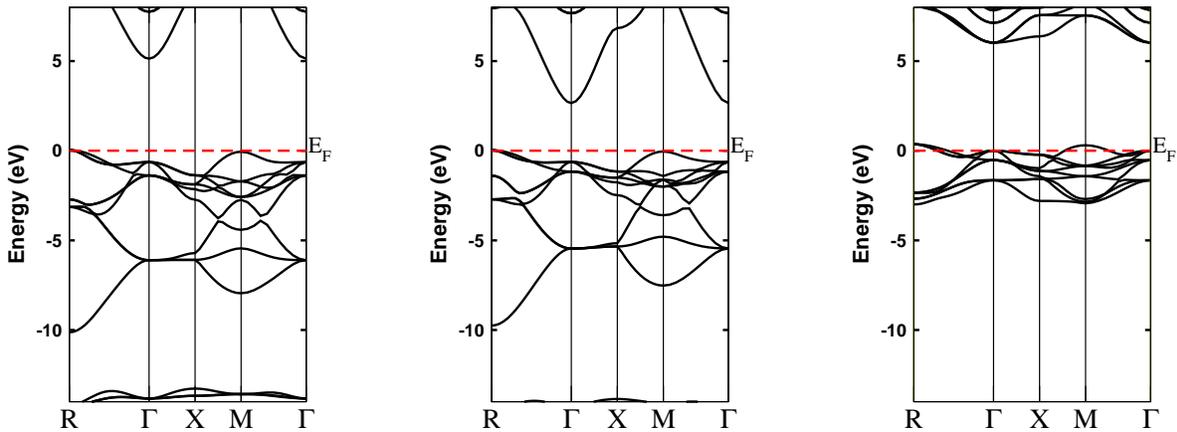

	\centering
	\includegraphics[scale=0.5]{Band_Si.eps} \hspace*{1cm}
	\includegraphics[scale=0.5]{Band_Ge.eps} \hspace*{1cm}
	\includegraphics[scale=0.5]{Band_Sc.eps}
	\caption{ Band Structure of cubic perovskites $SrSiO_3$, $SrGeO_3$ and $SrScO_3$ using nmBJ approach.} \label{Figure:3}
\end{figure}

We proceed now to make a detailed computation of electronic properties containing an investigation of PDOS, TDOS and a band structures. The major objective is to investigate the electronic behavior of these studied materials and also to understand the origin of the gap which appears for some treated compounds. Table \ref{Table:2} shows the band gap values of cubic $SrXO_{3}  \;(X=Sc,\;Ge,\;Si)$ using different exchange potentials such as: LDA, PBE-GGA, PBEsol-GGA, WC-GGA, mBJ, nmBJ and unmBJ.
\begin{table}[H]   	
	\begin{center}   		
		\begin{tabular}{ lllll }
			\hline
			Compounds                      & Exchange Functionals                 & Band gap (eV) \\ 
			\hline
			\multirow{7}{*}{SrGeO$_3$} 	& LDA                                  & 0.90    \\
			& PBE-GGA                              & 1.05    \\
			& PBEsol-GGA                           & 0.99    \\
			& WC-GGA                               & 0.97    \\
			& mBJ-GGA                              & 2.72    \\
			& unmBJ-GGA                            & 1.27    \\
			& nmBJ-GGA                             & 2.9    \\
			& Exp\cite{12}                         & 3.2    \\
			\hline
			\multirow{7}{*}{SrSiO$_3$}		 
			& LDA                                  & 3.52    \\
			& PBE-GGA                              & 2.73    \\
			& PBEsol-GGA                           & 3.18    \\
			& WC-GGA                               & 3.23    \\
			& mBJ-GGA                              & 5.18    \\
			& unmBJ-GGA                            & 3.97    \\
			& nmBJ-GGA                             & 5.37    \\
			& Others\cite{39}                      & (Insulator) \\           				
			\hline
			\multirow{7}{*}{SrScO$_3$}      
			& LDA                                  & metal    \\
			& PBE-GGA                              & metal    \\
			& PBEsol-GGA                           & metal   \\
			& WC-GGA                               & metal    \\
			& mBJ-GGA                              & metal   \\
			& unmBJ-GGA                            & metal    \\
			& nmBJ-GGA                             & metal   \\
			\hline
		\end{tabular}
	\end{center}
	\caption {Band gaps (in eV) of cubic 	$SrXO_3\;(X=Si,\;Ge,\;Sc)$ calculated using different exchange-correlation functionals.}
	\label{Table:2}
\end{table}
Actually, The LDA and GGA approximations were observed to have self-interaction exchange problems which give an underestimate of the band gap values \cite{40}. To solve these problems, we used the nmBJ potential which gives an accurate band gap results which are in good agreement with experimental studies \cite{29}. The reliable description of the fundamental band gap comes from the fact that this approach is based on a model potential like an orbital-dependent corrections which is implemented in the form of the functional used for the short-range part of the exchange potential \cite{28,41}.\\
Figure \ref{Figure:3}  shows the band structure of different studied perovskites according to $R \rightarrow \Gamma \rightarrow X \rightarrow M \rightarrow \Gamma$ path with high symmetry in the Brillouin zone. The Potential used as explained previously gave the most improved values of the gap. The minimum of $CB$ is at the $ \Gamma$ point and the maximum of $VB$ is at $M$ point for all studied compound which appeared in $\Gamma\rightarrow M$ transition. The electrons displacement which generate the movement of bands around $E_{f}$ especially at $ \Gamma$ are the main responsible of the electronic Behaviors presented for $X= Si,\;Sc,\;Ge$. An indirect band gap that is equal to $5\; eV$ for $SrSiO_3$, means that it behaves as an insulator. However, an interesting gap energy equal to $2.9\; eV$ appears for $SrGeO_3$ which presents an indirect band gap as well as p-type aspect. The substitution of $B$ atom by $Ge$-atom makes them a semi-conductor materials. The $SrScO_3$ shows a metal behavior with an overlap near Fermi level. We would like to stress that our results of the band structure of $SrGeO_3$ and $SrSiO_3$ show the same relevant features compared with $SrTiO_3$ where the band gap values are different \cite{42}. The obtained results of band gap energies are in good agreement with experiment and theoretical works \cite{12,19,42}.
\vspace{1cm}

\begin{figure}[H]
	\centering
	\includegraphics[scale=0.28]{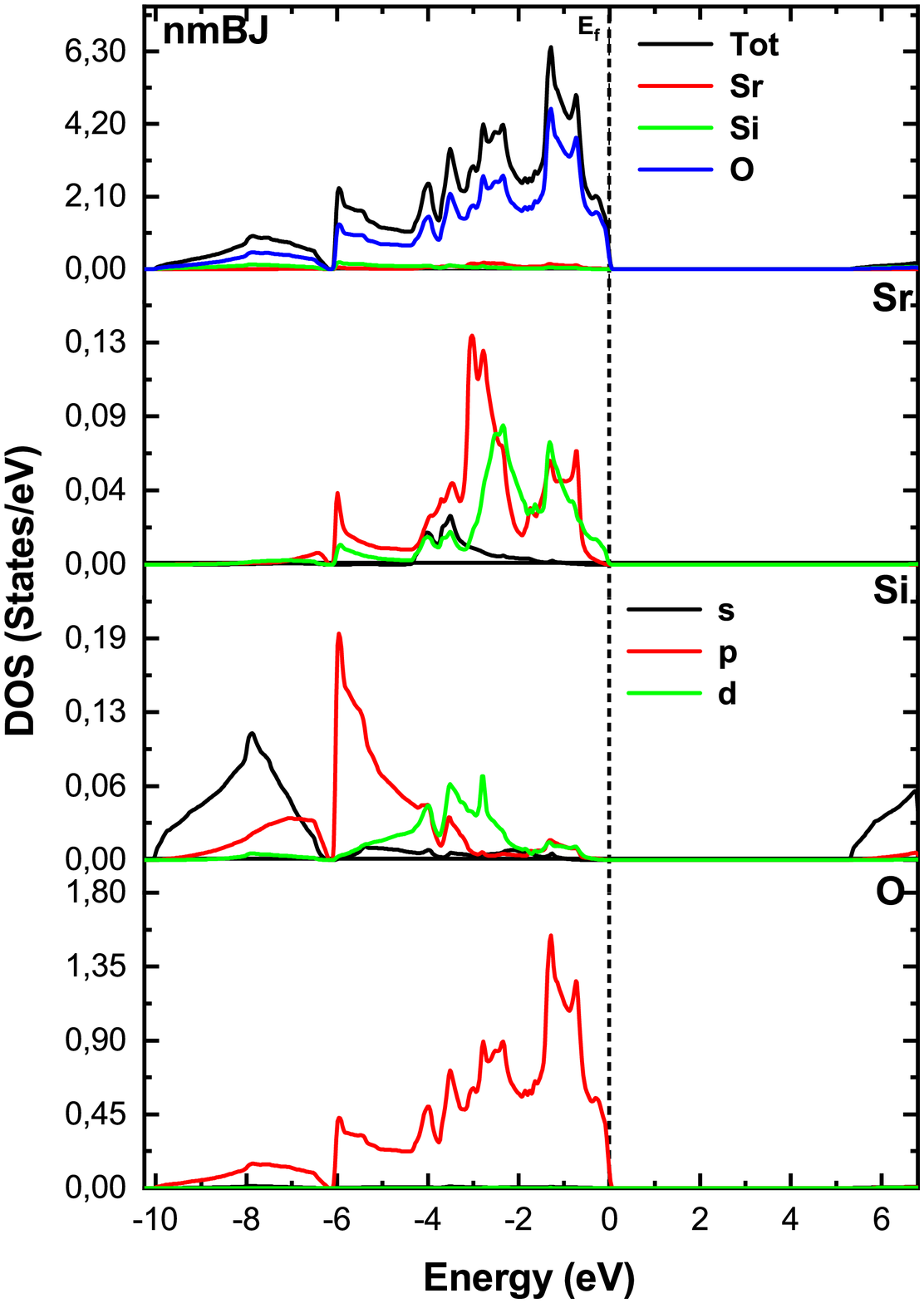} \hspace*{-0.4cm}
	\includegraphics[scale=0.28]{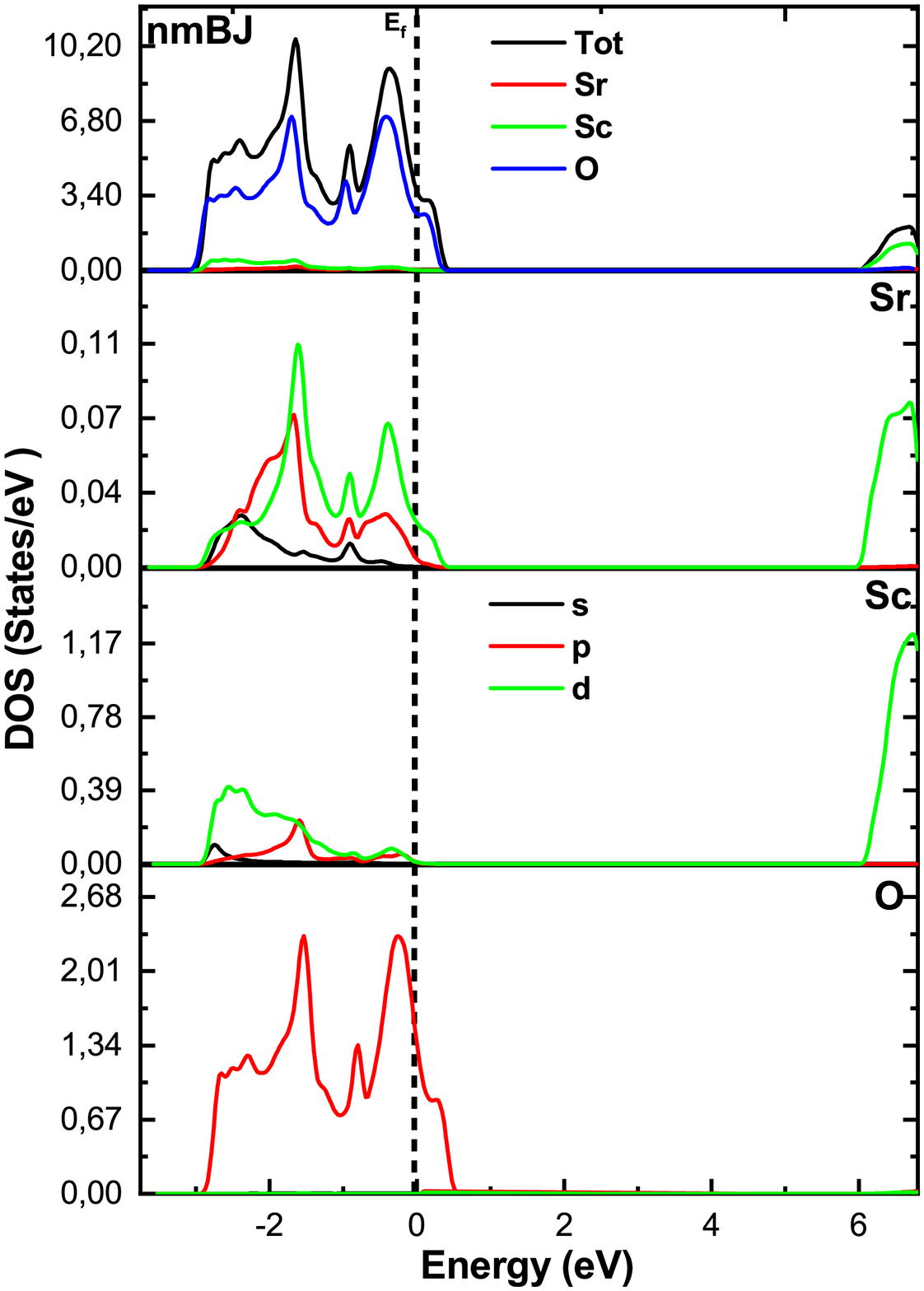} \hspace*{-0.4cm}
	\includegraphics[scale=0.28]{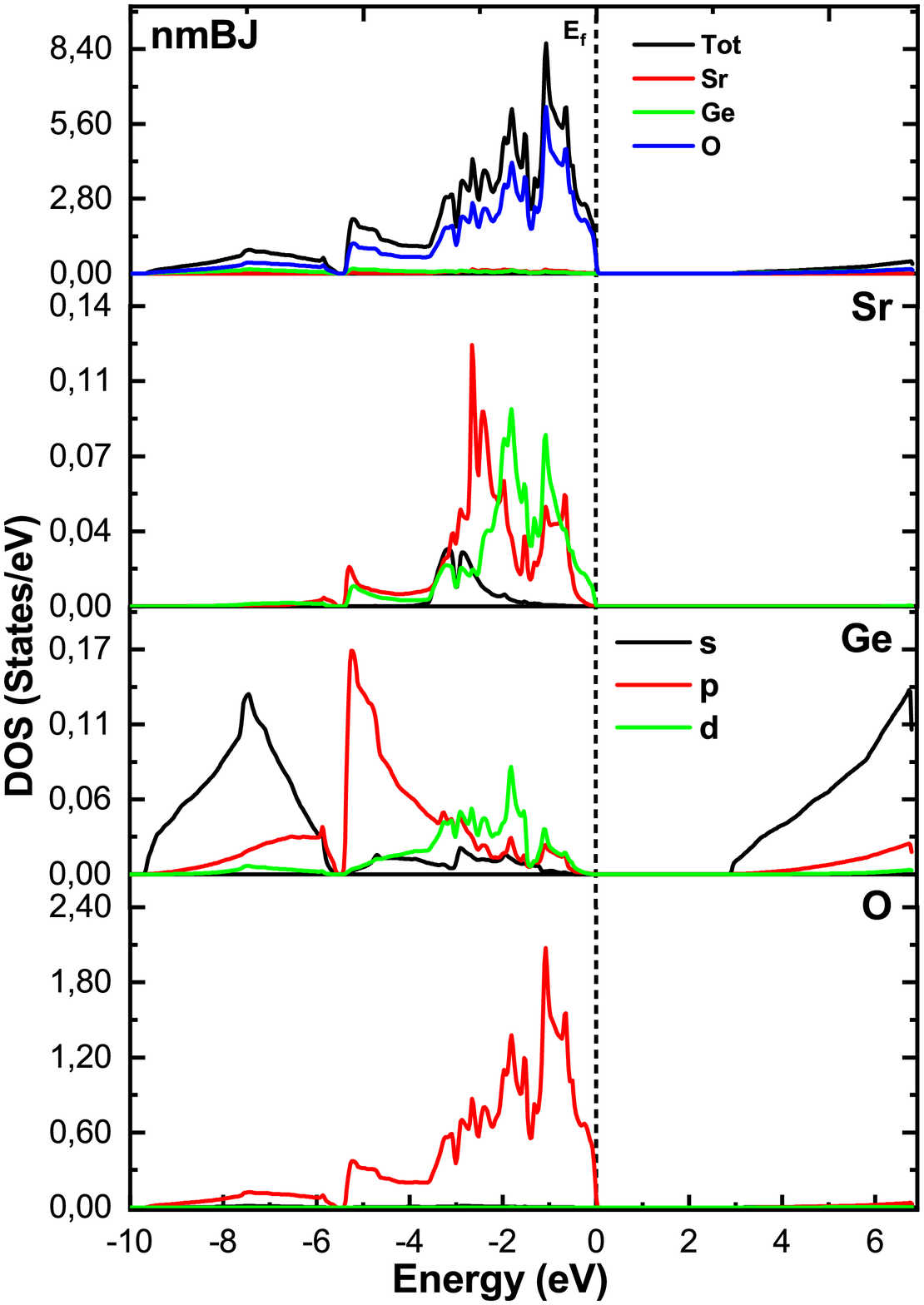}
	\caption{The partial density (PDOS) and total density of states (TDOS) of cubic perovskites $SrSiO_3$, $SrGeO_3$ and $SrScO_3$.}
	\label{Figure:4}
\end{figure}

To understand the origin of these behaviors, especially for fundamental band gap found, we proceed to study their partial density (PDOS) and total density of states (TDOS) in the energy range of $(E_{f} -10\; eV)$ to $(E_{f} + 6\; eV)$ in order to analyze the contribution of each atoms in the electronic configuration of the system as shown in figure \ref{Figure:4}.\\
Different energy levels give an appearance of different energy regions with different intra-bands around the Fermi level. In the case of $SrSiO_3$ compound (see figure \ref{Figure:4}), TDOS proves our previous comment about opening gap energy which is absolutely due to the lack of states energy around Fermi in the conduction band. At high energy $6\;eV$, the peaks showed is due to the contribution of Oxygen. The valence band is shown heavily near to Ef and mainly due to the occurrence of an intense intra-band between $-6 \;eV$ and $0\; eV$ constituted by $p$ and $d$ orbitals in $Sr$ and $Si$ atoms which gives $s-p$ hybridization. The strong electron transitions of this structure is between $Si$ and $Sr$ orbitals. Strontium atoms in $SrXO_3$ lattice are presented generally as $Sr^{2+}$ cations. The coupling of $d$ orbitals of $Sc$ and $p$ of Oxygen in case of $SrScO_3$, can appears a hybridization which can leads to the presence of covalent bonding between $Sc$ and $O$, this result is also shown for the case of $X= Ti$ \cite{43}. The $p$ orbitals of Oxygen with strong electron occupations and weak hybridization of $p$ and $d$ orbitals of $Sr$ that overlap around Fermi levels are the main responsible of the metallic aspect which appeared in this compound. Finally, $s$ of $Ge$ and $p$ orbital of Oxygen show the gap energy of this phase with an observed hybridization due to $p$ and $d$ orbitals of $Ge$ and $Sr$.
\subsection{Optical properties}
\subsection{Optical properties}

\begin{figure}[h!]
	\centering
	\includegraphics[scale=0.4]{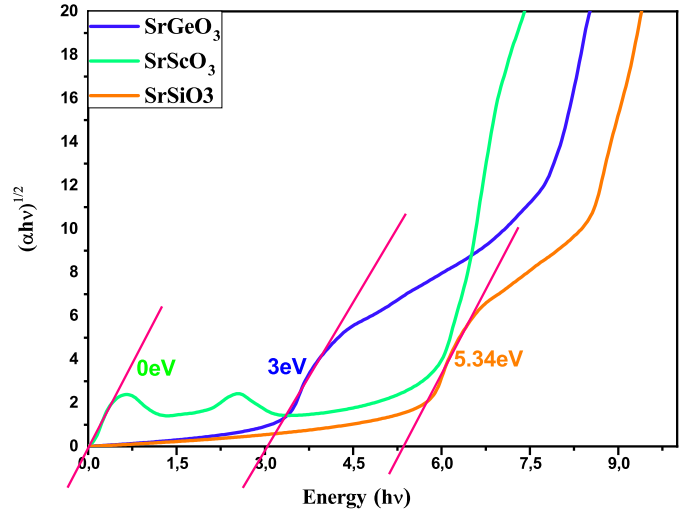}\hspace*{0.25cm}
	\caption{Optical band gap of $SrXO_3\; (X=Si,\; Ge,\; Sc)$}
	\label{Figure:5}
\end{figure}

\begin{figure}[H]
	\centering
	\includegraphics[scale=0.4]{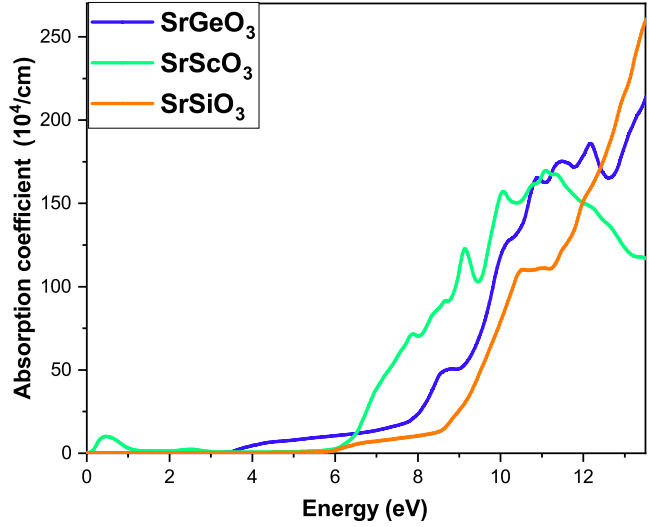}
	\caption{Absorption curves of the studied compounds}
	\label{Figure:6}
\end{figure}

\begin{figure}[H]
	\centering
	\includegraphics[scale=0.4]{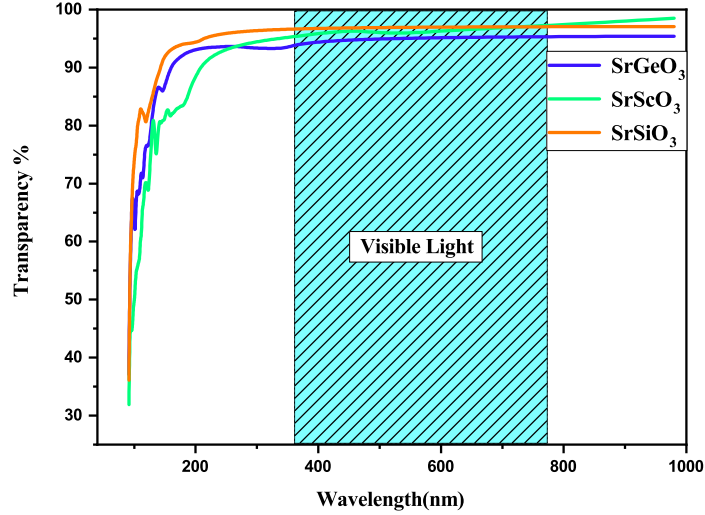}
	\caption{ Transmittance behavior of $SrXO_3$ with $X= Ge,\;Sc,\;Si$}
	\label{Figure:7}
\end{figure}

\begin{figure}[h!]
	\centering
	\includegraphics[scale=0.4]{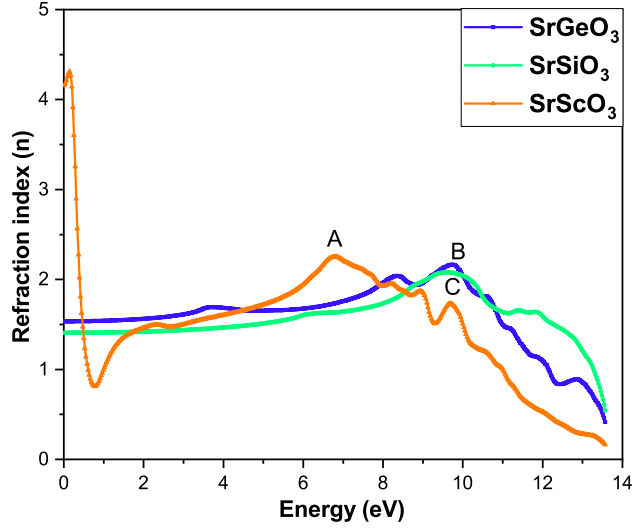}
	\caption{ Refraction index depending on energy variation}
	\label{Figure:8}
\end{figure}

To understand the optical behavior of these compounds, we proceed to study absorption, transparency and the refraction index which are very significant as a basic analysis in this issues. We started by the verification of the band gap obtained in the electronic investigation based on the following equation:
\begin{equation}
	(\alpha h \nu)^{k}= A(h\nu - E_g), \label{2}
\end{equation} 	
where $k$ is a constant which stands for the nature of the studied gaps. In our case, this index is equal to $ 1/2 $ since the gaps obtained are due to indirect transitions of electrons between the two bands. $A$ is also considered as constant which essentially depends on the possible transitions. Figure \ref{Figure:5}  depicts the variation of  $(\alpha h \nu)^{1/2}$ versus $h\nu$.
The optical band gaps obtained by this extrapolation are  $0\;eV$, $3\;eV$ and $5.34\;eV$ for $SrScO_3$, $SrGeO_3$ and $SrSiO_3$, respectively. These results are in good agreement with the electronic study which are $0\;eV,\;2.9\;eV$ and $5.18\;eV$ for $X= Sc,\;Ge$, and $Si$, respectively.
For more in-depth details regarding the optical analysis of these materials, we go to the absorption intensity which explains the ability of the particles  to absorb light especially in the visible light region. We focused on visible light since the objective of this work is to know the capacity of these materials to be  exploited such as electrodes in solar applications  or in the manufacturing of transparent sensors which is based on light detection effect. Nevertheless, these electrodes have been proven in the previous paragraph that they present a $p$-type in the transitions of the electrons where holes are the charge carriers responsible for a possible optical conductivity. This conduction type is considered as an enormous quality leads to an enhanced and exploitable optical behavior. The following curves in figure \ref{Figure:6} show the absorption of these materials.
From this figure, it is seen that below $300\;nm$, especially in ultraviolet-C, ($120-200\;nm$), $SrScO_3$ has low absorption and practically negligible in the area of visible light for all systems. This result will be checked by the calculation of the transparency, as shown in the figure \ref{Figure:7}. The transmittance shows a stable intensity from $300\;nm$ which means that $Sc$, $Si$ and $Ge$ have a desirable effect on the transparency. The averages of the transparency noted vary between $94\%$ and $97\%$ which is very usable in optoelectronics field. The interactions and interferences between light, in the visible range and matter, are described primarily by the laws of electromagnetism. The optical transparency of a material is distinguished by the non-absorption of energy in the matter. This property must be visualized for each wavelength spectrum. These studied elements present an important isotropy, since the light which penetrates them have not a privileged direction. When dealing with solid compound, absorption corresponds to the phenomenon of transformation and conversion of incident light energy into thermal energy or other form. 
Figure \ref{Figure:8} illustrates the static dielectric constant $\varepsilon_1(0)$ which is picked up at the equilibrium lattice constant and is about $2.34,\;1.96$ and $17.31$ for $SrGeO_3$, $SrSiO_3$ and $SrSCO_3$, respectively. Therefore, the extracted refractive index  $n(0)=\sqrt{\varepsilon_1(0)} $ is given in table  \ref{Table:3}.
\begin{table}[h!!]
	\begin{center}
		\begin{tabular}{ lllll }
			\hline
			Compounds		                  & $n(0)=\sqrt{\varepsilon_{1}(0)}$                  & $\varepsilon_{1}(0)$ \\
			\hline
			\\
			SrGeO$_3$                          & 1.53                                             & 2.34    \\
			\\
			SrSiO$_3$                          & 1.40                                             & 1.96    \\
			\\
			SrScO$_3$                          & 4.16                                             & 17.31   \\
			\hline
		\end{tabular}
	\end{center}
	\caption { Static refractive index and static dielectric constant of $SrSiO_3$, $SrGeO_3$, $SrScO_3$.}
	\label{Table:3}
\end{table}
Figure \ref{Figure:8} also shows the variation and modulation of the refractive index as a function of the energy. It gives various maximum picks at $A$ ($E=6.7\;eV$,$n=2,25$) for $Sc$, $B$($E=9,7\;eV$,$n=2,15$) for $Ge$ and $C$($E=9,7\;eV$,$n=2,04$) for $Si$. These picks are explained respectively by the low transmittance between $206\;nm$ and $124\;nm$ for all treated compounds. Below $4\;eV$, the index is relatively stable and influences directly the transparency aspect.

\subsection{Thermoelectric properties and Phonon dispersions}
\subsubsection{Thermoelectric properties}
The thermoelectric materials (TE) have a major interest because of their ability to convert thermal heat from industrial processes into usable electricity\cite{44, 45}. Thermoelectric and transport properties have been investigated using the semi-classical Boltzmann theory which is implemented in the BoltzTraP code. 
In addition, the electrical conductivity depending on relaxation time $\sigma/\tau$ is studied using nmBJ exchange potential. Where, The seebeck coefficent,the electrical and thermal conductivities are given by:
\begin{equation}
	\begin{aligned}
		S_{\alpha\beta} (T, \mu) &= \dfrac{1}{eT\sigma_{\alpha\beta} (T, \mu)\Omega} \int \sigma_{\alpha\beta} (\varepsilon)(\varepsilon - \mu) \left[\dfrac{\partial f_{\mu}(T, \mu)}{\partial \varepsilon} \right] d\varepsilon,
	\end{aligned}
\end{equation}			
\begin{equation}
	\begin{aligned}
		\sigma_{\alpha\beta} (T, \mu) &= \dfrac{1}{\Omega} \int \sigma_{\alpha\beta} (\varepsilon) \left[\dfrac{\partial f_{\mu}(T, \mu)}{\partial \varepsilon} \right] d\varepsilon,
	\end{aligned}
\end{equation}
\begin{equation}
	\begin{aligned}
		\kappa_{\alpha\beta} (T, \mu) &= \dfrac{1}{e^{2}T\Omega} \int \sigma_{\alpha\beta} (\varepsilon)(\varepsilon - \mu)^{2} \left[\dfrac{\partial f_{\mu}(T, \mu)}{\partial \varepsilon} \right] d\varepsilon,
	\end{aligned}
\end{equation}	
With $\Omega$ is the supercell volume, $\mu$ indicates the chemical potential  and $\alpha$, $\beta$ correspond on the tensor indicators. The $f_\mu$ is the Fermi distribution.\\

\begin{figure}[h]
	\centering
	\includegraphics[scale=0.6]{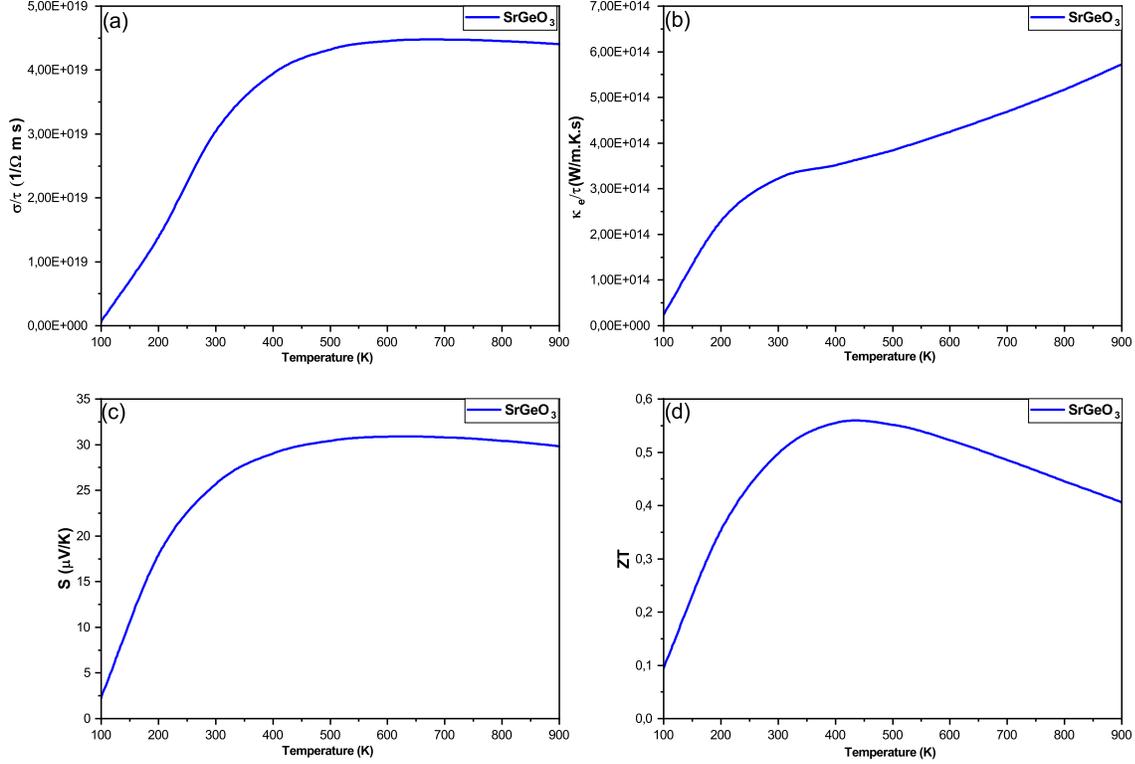}
	\caption{ Thermoelectric properties depending en temperature: (a) Electrical conductivity, (b) Thermal conductivity, (c) Seebeck coefficient, (d) Figure of merit for $SrGeO_{3} $}
	\label{Figure:9}
\end{figure}

The performance of a TE materials is measured by figure of merit parameter $ZT$ ,given by $ZT = {S^{2}\sigma T}/{(\kappa_{lat} + \kappa_{ele})}$, where $\kappa_{lat} \,and\, \kappa_{ele}$ are lattice thermal conductivity, and electronic thermal conductivity, respectively \cite{46}. Due to limitation of the BoltzTraP code, the $ZT$ values obtained are  slightly overestimated because of the phonon contribution to the thermal conductivity which is ignored. The BoltzTraP code treats the calculation of lattice thermal conductivity contributed by phonons as a constant and gives the electronic part of the thermal conductivity, the conductivities (electronic and thermal) correspond proportionally on the relaxation time $\tau$. Those values are then adjusted by a multiplication with relaxation time constant to find the final transport properties \cite{46}. The expression of the $ZT$ merit factor  summarises the difficulty of optimising the transport properties of a thermoelectric material. Intuitively, it seems difficult for a material to simultaneously possess low thermal and high electrical conductivities. Ideally, a good thermoelectric material must have both, the electrical conductivity of a metal and the thermal conductivity of an insulator, which is rarely obtained. This duality can be seen in  some semiconductor class\cite{46}.  Therefore, we are going to focus on  compound which has semiconductor behavior in our studies ($SrGeO_3$).\\
In this part, we studied the thermoelectric properties of $SrGeO_3$ compounds using the semi-classical theory of the Boltzmann. The transport properties such as  the Seebeck coefficient $(S)$, the electrical conductivity relative to relaxation time $(\sigma/\tau )$, the thermal conductivity relative to relaxation time $(\kappa/\tau)$ and the figure of merite parameter $(ZT)$ versus the variable temperatures are performed and depicted in Figure \ref{Figure:9}. \\
Figure \ref{Figure:9}(a) shows this variation depending on temperature from $100\;K$ to $900\;K$. The minimum value of electric conductivity ($\sigma / \tau$) for the $SrGeO_3$ compound is equal to  $0.62\times 10^{18}\, 1/(\Omega m s)$  at $T = 100\;K$ and its maximum value is equal $4.5\times 10^{19}\, 1/(\Omega m s)$  at $T = 600\;K$, this indicates that $SrGeO_3$ is a semiconductor material which does not require a strong excitation and thermal agitation to accelerate the mobility of the charge carriers.  The $\sigma/\tau$ presents a direct proportionality with temperature which reveals that there are greater concentrations of carriers indicating that the electrons are thermally excited. The figure \ref{Figure:9}(b) shows Thermal conductivity of the material which reachs maximal values at 800 K, and the $\kappa_{e} $ value  increases with rise in temperature, this is caused by the increasing of electron energies with temperature. However, $\kappa_{e} $ presents a small values at room temperature which equals to $ 3$ $W/m.K$. The calculated Seebeck coefficient regarding the temperature is shown in the figure \ref{Figure:9}(c) which has a positive sign, this sign shows that the studied compound is p-type semiconductor. Furthermore, the dominated charge carriers responsible for conduction in the treated compound are holes. \\
The figure \ref{Figure:9}(d) shows the variation of the merit parameter for $SrGeO_3$ compound with as a function of temperature. It has been observed that $ZT$ improved with increasing in temperature and obtain a maximum value at $380\;K$. However, at room temperature, the value of $ZT$ equals to $0.5$. This value of $ZT$ parameter is appropriate for potential thermoelectric material. Thus, it seems that $SrGeO_3$ behaves as a good thermoelectric material for future energy applications.    
\subsubsection{Phonon dispersions}
\begin{figure}[H]
	\centering
	
	\includegraphics[scale=0.262]{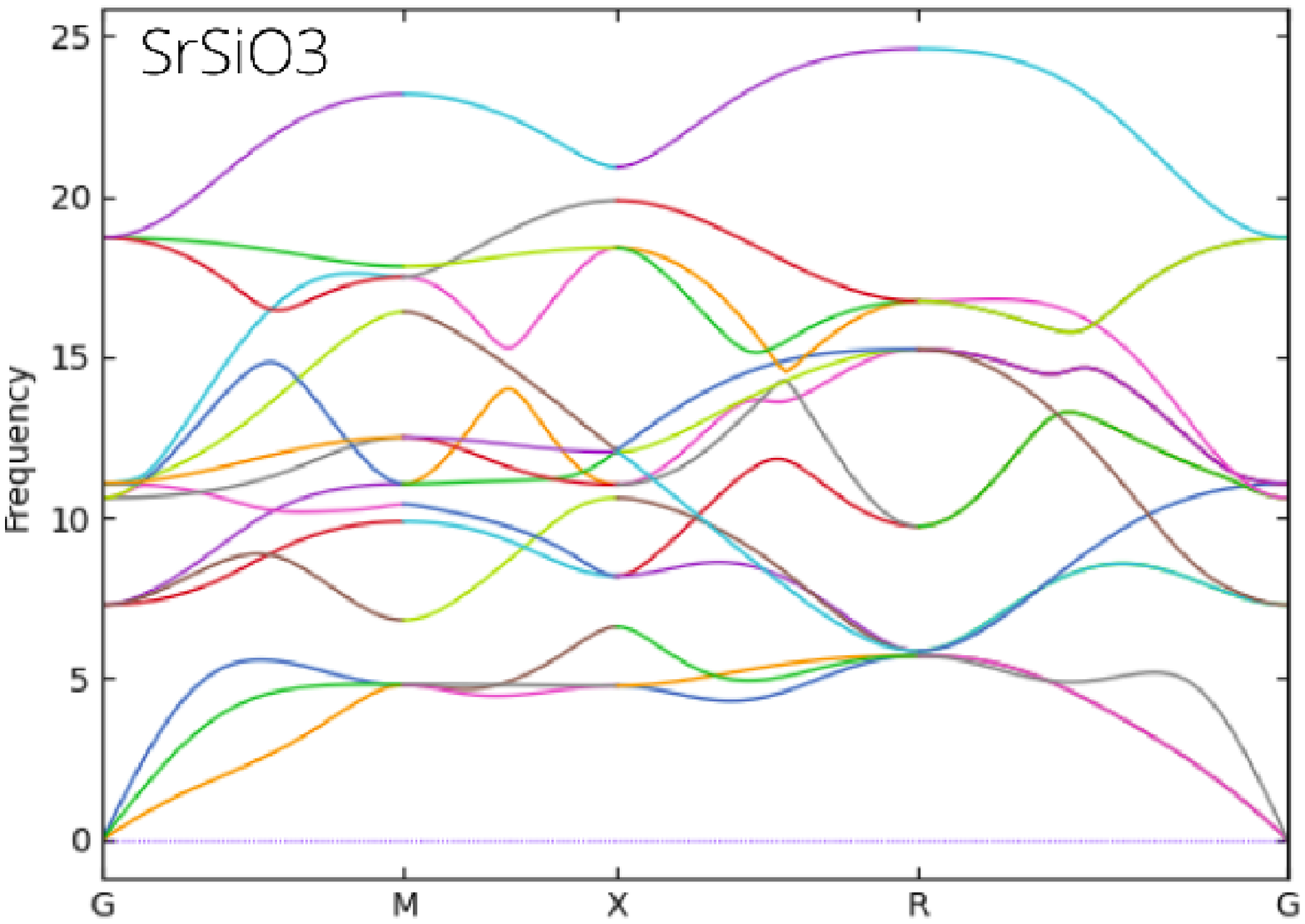} 
	\includegraphics[scale=0.252]{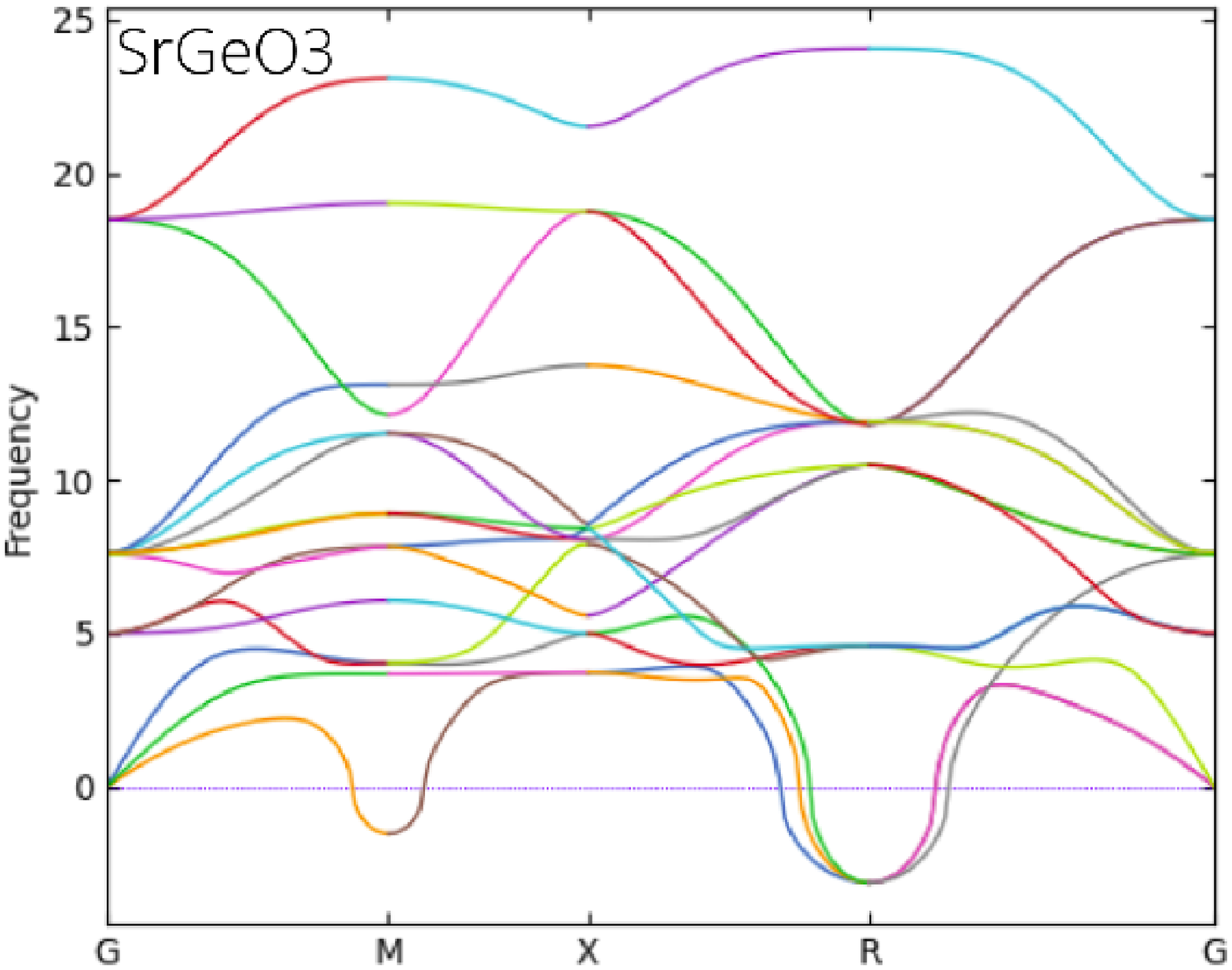} 
	\includegraphics[scale=0.26]{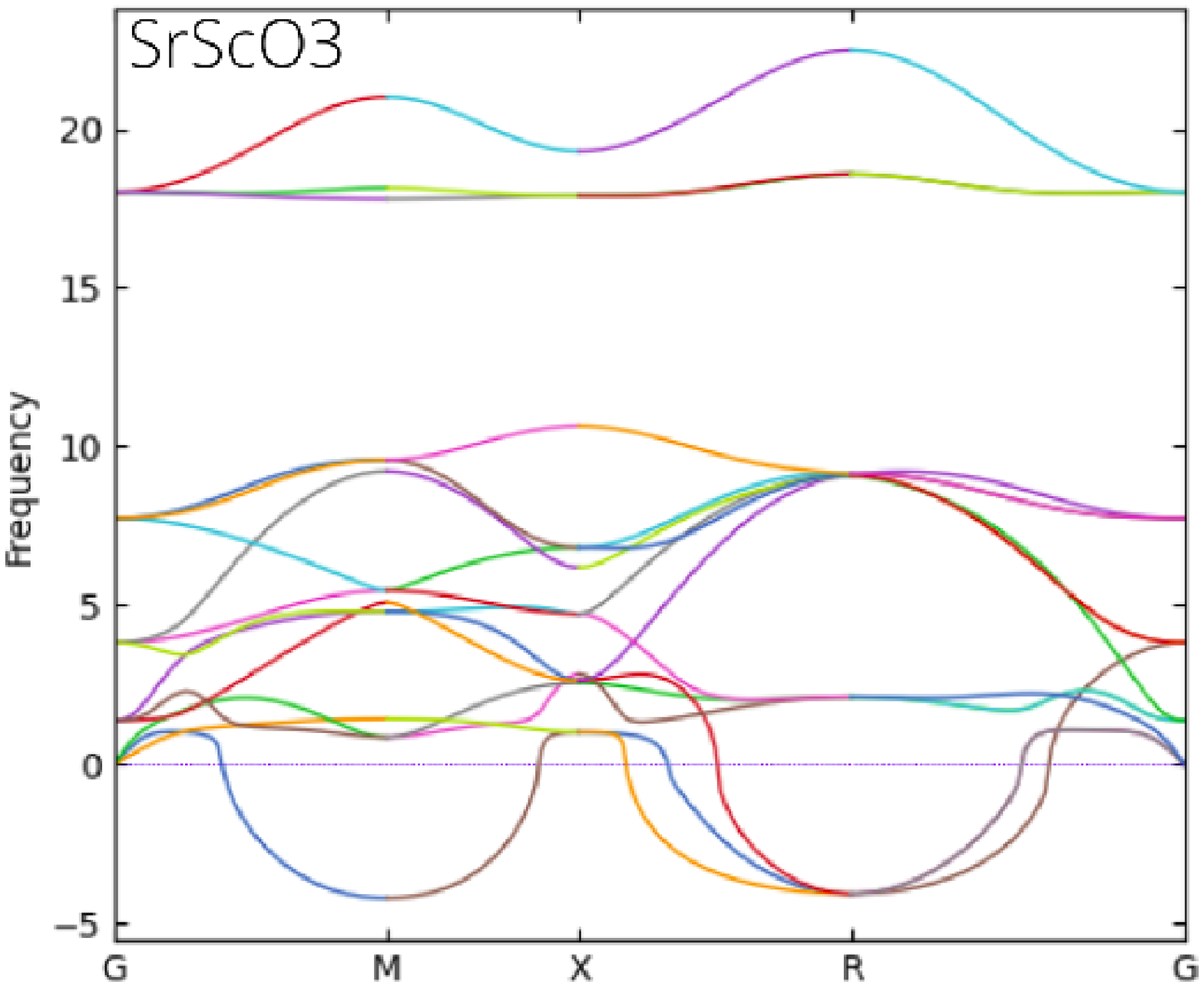}
	\caption{ Phonon dispersions of cubic perovskites $SrSiO_3$, $SrGeO_3$ and $SrScO_3$.} \label{Figure:10}
\end{figure}

The phonons are calculated using Phonopy package \cite{47} from  the potential energy surface around a stationary point, the matrix of force constants is given by:
\begin{equation}
	D_{i\beta,i^{'}\beta^{'}}(\textbf{R}_{p},\textbf{R}_{p'})=\dfrac{\partial^{2}E}{\partial u_{p\beta i}\partial u_{p' \beta' i'}}
\end{equation}
In Cartesian direction, $u_{p\beta i}$ is the displacement term of atom $\beta$ taken in the supercell at $R_{p}$, the term of $E$ is considered as the potential energy surface in which the nuclei displaces. This parameter gives the curvatures in one point taken as reference, derivate in second order energy in $(x,\;y,\;z)$ directions. Therefore using the following dynamical matrix, the phonons can found at every point of the Brillouin zone:
\begin{equation}
	D_{i\beta,i'\beta'}(\textbf{q})=\dfrac{1}{N_{p}\sqrt{m_{\beta} m_{\beta'}}} \sum_{\textbf{R}_{p},\textbf{R}_{p'}}D_{i\beta,i'\beta'}(\textbf{R}_{p},\textbf{R}_{p'})e^{i\textbf{q}(\textbf{R}_{p}-\textbf{R}_{p'})}
\end{equation}
When $m_{\beta}$ is the mass of atom $\beta$ and $N_p$ is the number of cells in the supercell structure studied. Physically phonons are the bosonic quasi-particles associated by $(q,\nu)$ as quantum numbers, eigenvalues $\omega^{2}_{q\nu}$ and eigenvectors $V_{q\nu}$, $i\beta$ are found by diagonalizing the previous dynamical matrix. Figure \ref{Figure:10} shows the phonon dispersions of the studied systems.
For $SrScO_3$ the structure is observed dynamically unstable since the eigenvalues shown (Fig10) are negative in the presented curvatures, which correspond to imaginary phonon frequencies. The energy decrease in the PES direction. For $X=Ge,\; Si$ most eigenvalues $\omega^{2}_{q\nu}$ of the dynamical matrix  are  shown  relatively positive, the phonon frequencies are then all real, the structures are considered stable dynamically. At $R$ point, the $SrGeO_3$  presents an imaginary frequency which can be explained by a critical temperature leads to a variation of the  energy surface at the Brillouin zone boundaries \cite{12}.
	\newpage
	\section{Conclusion}
	
	The structural, electronic, optical and thermoelectric properties of cubic perovskite oxides $SrXO_3\;(X = Si,\; Ge,\; Sc)$ have been studied using the first-principle calculations implemented in WIEN2K code. The investigation and optimization of the lattice constants are in good agreement with experimental results. Electronic properties demonstrate the insulator characteristic of $SrSiO_3$ compound with the indirect band gap, semiconductor for $SrGeO_3$ with ($E_g=2.9\;eV$ ) and the metallic behavior of $SrScO_3$ compounds. The transmittance shows a stable behavior from $300$ $nm$ with an average of  $94\%$ and $97\%$. A significant conductivity and good figure of merit value for $X$ $=$ $Ge$ are shown at room temperature $T=298,15$ $K$. The phonon dispersion was made to verify the stability of these perovskites, all results found show that the presence of Ge and Si leads to  a dynamic stability of the structure. These results make these materials as a promoter candidates in optoelectronic applications (detection sensors) or even as a strong alternatives to TCOs with the quality of a p-type conduction.
	
\end{document}